\documentstyle[12pt,epsf,amssymb]{article}

\begin{document}

\baselineskip=18.8pt plus 0.2pt minus 0.1pt

\def\CR{\nonumber \\}
\def\lp{l_P}
\def\pt{\partial}
\def\be{\begin{equation}}
\def\ee{\end{equation}}
\def\bea{\begin{eqnarray}}
\def\bee{\end{eqnarray}}
\def\eq#1{(\ref{#1})}

\def\R{{\mathbb R}}
\def\tr{{\rm tr}}
\def\ch{{\rm cosh}}
\def\sh{{\rm sinh}}
\def\del{\partial}
\def\la{\langle}
\def\ra{\rangle}
\def\dg{\int_0^{2\pi}d\phi \int_0^{2\pi}d\theta \int_1^{\infty}\frac{dx}{2}}

\begin{titlepage}
\title{
\hfill\parbox{4cm}
{\normalsize KUNS-1665 \\{\tt hep-th/0005178}}\\
\vspace{1cm}
Scalar field theories in a Lorentz-invariant three-dimensional noncommutative
space-time}
\author{
Shin'ichi {\sc Imai}\thanks{\tt imai@gauge.scphys.kyoto-u.ac.jp}~ 
and
Naoki {\sc Sasakura}\thanks{\tt sasakura@gauge.scphys.kyoto-u.ac.jp}
\\[7pt]
{\it Department of Physics, Kyoto University, Kyoto 606-8502, Japan}}
\date{\normalsize May, 2000}
\maketitle
\thispagestyle{empty}

\begin{abstract}
\normalsize
We discuss scalar quantum field theories in a
Lorentz-invariant three-dimensional noncommutative space-time.
We first analyse the one-loop diagrams of the two-point functions,
and show that the non-planar diagrams are finite and have
infrared singularities from the UV/IR mixing.
The scalar quantum field theories have the problem that the violation of
the momentum conservation from the non-planar diagrams
does not vanish even in the commutative limit.
A way to obtain an exact translational symmetry by introducing an infinite
number of tensor fields is proposed.
The translational symmetry transforms local fields into non-local
ones in general.
We also discuss an analogue of thermodynamics of free scalar field
theory in the noncommutative space-time.
\end{abstract}
\end{titlepage}

\noindent
\section{Introduction}
Quantum field theories in noncommutative space-times with 
the noncommutativity $[x^\mu,x^\nu]=i\theta^{\mu\nu}$  are interesting 
in pursuing the new possibilities of quantum field theories in
quantum space-times.  
Remarkably, this kind of noncommutativity between coordinates appears
in string theory, for instance, 
in the toroidal compactification of Matrix Theory \cite{con} and in
open string theory \cite{CH99}-\cite{SW99} with a B-field background.
Thus it may be expected that field theories in this kind of noncommutative
space-times are controllable in some way.
On the other hand, however, 
perturbative analyses show that these noncommutative
field theories have interesting but unusual 
behaviours \cite{minw}-\cite{mozo2}.
Infrared singularities in the correlation functions
were shown to appear even for massive theories \cite{minw}-\cite{raam}.
In the case with space-time noncommutativity with 
$\theta^{0i}\neq 0$ \cite{seisuss}-\cite{gomis},
the causality is violated in interesting ways \cite{seisuss}, 
and S-matrix does not satisfy unitarity constraints \cite{gomis}.
Thus, before we become able to handle such noncommutative quantum field
theories consistently, there still seems to remain a lot to learn about them.

An obvious direction to learn more would be to generalise the class of
noncommutative space-time. A constant background of $\theta^{\mu\nu}$
violates Lorentz symmetry in more than two-dimensional space-time.
Since Lorentz symmetry is one of the fundamental components in
the present theoretical physics, in this paper we consider a Lorentz-invariant
three-dimensional space-time with the noncommutativity
$[x^\mu,x^\nu]=2 i\l_P \varepsilon^{\mu\nu\rho}x_\rho$\footnote{A 
noncommutative space-time with this commutation relation is often
quoted a quantum sphere. For example, see \cite{haw}-\cite{wat}. 
In this paper, the noncommutative space-time is three-dimensional,
rather than a two-dimensional subspace in it. See also \cite{madore}.}.
Though this noncommutativity is different from that obtained
from the constant background of the two-form field in string theory, 
a similar kind of noncommutativity between more than two coordinates
appears on the boundary string of a membrane in M-theory with a
non-vanishing background of the three-form field \cite{berg,kawa}, and
also on a D2-brane in a non-constant two-form field background in
string theory \cite{ho}.
Another motivation comes from that this noncommutative space-time may
be regarded as a fuzzy space-time with a Lorentz-invariant space-time
uncertainty relation \cite{sasa2} derived from a gedanken experiment
in which only the general relativity and quantum mechanics
are used \cite{karo}-\cite{ahl}.
Concerning the questions whether there is any relationship between the
space-time uncertainty relation and quantum gravity or string theory 
\cite{yoneya}, 
as well as whether the properties obtained so far for the specific
noncommutative quantum field theories are general in other
noncommutative space-times,
it would be interesting to investigate quantum field theories
in that noncommutative three-dimensional space-time.
As for the latter question, since the time-coordinate is also
noncommutative in the space-time we consider, 
it should be the most radical case with the
above-mentioned violations of causality and unitarity
\cite{seisuss}-\cite{gomis}.
In this paper, we shall discuss only scalar field theories in the
noncommutative space-time.

The organisation of this paper is as follows.
In section \ref{sec:hilbert}, we discuss the group theoretical
structure of the one-particle Hilbert space of free scalar field theory.
In section \ref{sec:IntFeyn},
we define the action and derive the Feynman rules.
In section \ref{sec:1Loop}, we compute some one-loop diagrams.
The non-planar diagrams are shown to be finite and have infrared
singularities from the UV/IR mixing.
We encounter the feature that the violation of the momentum
conservation from the non-planar diagrams does not vanish even
in the commutative limit $\lp\rightarrow 0$.
In section \ref{sec:trans}, we propose a translationally symmetric
theory to remedy the defect.
In section \ref{sec:thermo}, we discuss a noncommutative analogue of the
thermodynamics of free scalar field theory and
compare the result with the qualitative estimation
given previously in \cite{sasa1}.
Section \ref{sec:sum} is devoted for the summary and discussions.

\section{One-particle Hilbert space}
\label{sec:hilbert}

In the paper \cite{sasa2}, one of the present authors
discussed the momentum space representation of the one-particle Hilbert
space of free scalar field theory in a Lorentz-invariant
three-dimensional noncommutative space-time.
There the noncommutativity of the space-time is motivated by a space-time
uncertainty relation derived from a certain gedanken experiment
\cite{sasa2}-\cite{sasa1}.
In this section we discuss the group theoretical structure of the
one-particle Hilbert space.

\subsection{Construction via $ISO(2,2)$ algebra}

The three-dimensional noncommutative space-time in \cite{sasa2} is represented
by the following $SO(1,2)$ Lorentz-invariant commutation relations
between the coordinates and momentum
operators\footnote{ 
We have rescaled the numerical constant of the algebra 
by a factor of 2 from that in the previous paper \cite{sasa2}.}: 
\begin{eqnarray}
  \label{NCPhaseSpace}
[\hat{x}^{\mu},\hat{x}^{\nu}] &=& 2i l_{P} 
\epsilon^{\mu\nu\rho} \hat{x}_{\rho},\cr
[\hat{p}^{\mu},\hat{p}^{\nu}] &=& 0, \cr 
[\hat{p}^{\mu},\hat{x}^{\nu} ] &=& -i \eta^{\mu\nu} \sqrt{ 1 
+ l_P^2 \hat{p}^2 } 
+ i l_P  \epsilon^{\mu\nu\rho} \hat{p}_\rho,
\end{eqnarray}
where the greek indices run through 0 to 2, and the
numerical constant $\lp$ should be in the order of Planck length.

Now let us consider the lie algebra of $ISO(2,2)$, 
\begin{eqnarray}
  \label{iso(2,2)}
[\hat{J}_{mn},\hat{J}_{kl}] &=& 
-i(\eta_{mk}\hat{J}_{nl} - \eta_{ml}\hat{J}_{nk}
- \eta_{nk}\hat{J}_{ml} + \eta_{nl}\hat{J}_{mk}),\cr
[\hat{J}_{mn},\ \hat{p}_{k}] &=& -i(\eta_{mk}\hat{p}_n 
- \eta_{nk}\hat{p}_m), \cr
[\ \hat{p}_m,\ \hat{p}_n] &=& 0,
\end{eqnarray}
where the roman indices run through $-1$ to 2,
and the signature is given by $\eta_{mn} = (-,-,+,+)$. 
By identifying
\bea
  \label{JtoX}
  \hat{x}_\mu &=& l_P(\hat{J}_{-1,\mu} - \frac{1}{2}
  \epsilon_{\mu}{}^{\alpha\beta} \hat{J}_{\alpha\beta}) ,\cr
\hat{p}_\mu&=&\hat{p}_{m=\mu},
\bee
and imposing a constraint
\begin{equation}
\label{Constraint}
   1 + l_P^2 \hat{p}^m \hat{p}_m = 0,  
\end{equation}
we can easily show that
the commutation relations \eq{NCPhaseSpace} can be derived from
\eq{iso(2,2)}.
The momentum square $\hat{p}^m \hat{p}_m$ is one of the Casimir
operators of the algebra \eq{iso(2,2)}, 
and hence we may consistently impose the constraint \eq{Constraint}.
The remaining three independent generators in \eq{iso(2,2)} 
are now the $SO(1,2)$ Lorentz generators of the noncommutative 
space-time \eq{NCPhaseSpace}. 

The representation of the algebra (\ref{NCPhaseSpace}) is
obtained from that of $ISO(2,2)$, which is given by
$\hat{J}_{mn} = i(p_m \frac{\del}{\del p^n} - p_n \frac{\del}{\del p
^m})$
in the momentum space representation.
In order to impose the constraint (\ref{Constraint}), the
following ``polar'' coordinate is convenient:
\begin{eqnarray}
  \label{PolarCoordinate}
  p_{-1} &=& r \ch \chi \cos \theta, \cr
  p_0    &=& r \ch \chi \sin \theta, \cr
  p_1    &=& r \sh \chi \cos \phi, \cr
  p_2    &=& r \sh \chi \sin \phi.
\end{eqnarray}
This coordinate is only valid in the neighbourhood of the hyperboloid
$1 + l_P^2 p^m p_m =0$, but this is enough
for our purposes.
Since $\hat{J}_{mn} p^l p_l = 0$ and $\hat{J}_{mn}$ do not contain
$r$-derivative, we can restrict the representation space
to the functions on the hyperboloid, i.e. the functions
depending only on $\chi, \theta, \phi$.
Thus the natural inner-product that makes $\hat{J}_{mn},\hat{p}_m$ hermite
in the restricted representation space is given by
\begin{eqnarray}
  \label{measure4}
  \la \Phi_1 | \Phi_2 \ra =
2 \int d^4p\ \delta(1+l_P^2 p^m p_m) \Phi_1^*(p) \Phi_2(p).
\end{eqnarray}
After integrating over $p_{-1}$,
this inner product is identical to that obtained previously in \cite{sasa2}.

The constraint \eq{Constraint} shows
that the mass square $-p^\mu p_\mu$ in the noncommutative space-time
should have an upper bound $-p^\mu p_\mu \leq 1/l_P^2$, and also
that, coming from the choices of the sign of $p_{-1}$, there exists two-fold
degeneracy in the momentum space of the noncommutative space-time.
These features agree with the results obtained in \cite{sasa2}.

\subsection{$SL(2,R)$ structure of the momentum space}
\label{sec:SL2Rstr}

In the previous subsection, we have shown that the momentum 
space is the hyperboloid,
$1 + l_P^2 p^m p_m =0$.
This hyperboloid can be mapped to the group manifold
$SL(2,R)$ as follows. Let us define the matrices
\begin{eqnarray}
  \label{2Rep}
  {\bold x}^0 &=& \lp \sigma^2 = l_P \left( \matrix{ 0 & -i \cr i & 0 }
  \right) ,\cr
  {\bold x}^1 &=& \lp i \sigma^3 = l_P \left( \matrix{ i & 0 \cr 0 & -i }
  \right) ,\cr
  {\bold x}^2 &=& \lp i \sigma^1 = l_P \left( \matrix{ 0 & i \cr i & 0 }
  \right) ,
\end{eqnarray}
which satisfy
\begin{eqnarray}
  \label{relation}
{\bold x}^\mu {\bold x}^\nu = - l_P^2 \eta^{\mu\nu} + i l_P
\epsilon^{\mu\nu\lambda}{\bold x}_\lambda.
\end{eqnarray}
Then the map between the hyperboloid and the group elements of
$SL(2,R)$ is defined by
\begin{eqnarray}
  \label{GroupElement}
  g = l_P \left( \matrix{ p_{-1}-p_1 & p_0 - p_2 \cr - p_0 - p_2 & p_{-1}
  + p_1 } \right) = l_P p_{-1} + i{\bold x}^\mu p_\mu = e^{ik{\bold x}}.
\end{eqnarray}
By a direct computation, we can verify
$p_\mu = k_\mu \sh(l_P \sqrt{k^2})/l_P\sqrt{k^2}$.
As discussed in \cite{sasa2},
this relation is identical to the relation between $k_\mu$ and the
$\hat{p}_\mu$-eigenvalue of the state $e^{-ik\hat{x}}|0\ra$,
where $|0\ra$ denotes the momentum-zero eigen-state with $p_{-1}=1$.
Thus the one-particle Hilbert space can be parameterised
by the group manifold $SL(2,R)$:
\begin{eqnarray}
  \label{state}
  |p(g)\ra = e^{i k(g) \hat{x}}|0\ra,
\end{eqnarray}
where $p_m(g)$ and $k_\mu(g)$ are defined by \eq{GroupElement}.
This parameterisation with $SL(2,R)$ elements is superior to the
parameterisation by the
space-time momentum $p_\mu$, because distinct choices of the signs of $p_{-1}$
correspond to distinct group elements and we do not need to worry
about the two-fold degeneracy in the momentum space representation.

For later convenience, we collect some useful formulae in the followings.
{}From (\ref{relation}) and \eq{GroupElement},
we can show that the group multiplication can be
expressed in terms of $p_m$ by
\begin{eqnarray}
\label{Multiplication}
p(gh)_{-1} &=& \lp (p(g)_{-1}p(h)_{-1} + p(g)^{\alpha} p(h)_\alpha), \cr
p(gh)_\mu &=& \lp (p(g)_{-1}p(h)_\mu + p(g)_\mu p(h)_{-1} -
\epsilon_\mu{}^{\alpha\beta} p(g)_\alpha p(h)_\beta).
\end{eqnarray}
Note that $\tr(g) = 2 p(g)_{-1}= \pm2\sqrt{\lp^{-2}+p_\mu(g)p^\mu(g)}$.
Since an adjoint action keeps the trace invariant, it keeps
$p_\mu p^\mu$ invariant.
Hence the adjoint action of $SL(2,R)$ corresponds to the $SO(1,2)$
Lorentz transformation of the space-time.
Under the adjoint action, $p(h)_m$ transforms as
\begin{eqnarray}
  \label{LorentzTrnsf}
  p(g^{-1}hg)_{-1} &=& p(h)_{-1}, \cr
  p(g^{-1}hg)_\mu &=& p(h)_\mu (1+2a^\alpha a_\alpha)
 -2 p(h)^\alpha a_\alpha a_\mu
-2\epsilon_{\mu}{}^{\alpha\beta}p(h)_{\alpha}a_\beta a_{-1},
\end{eqnarray}
where $a_m = \lp p(g)_m$.

After a short computation, we can verify that the inner product defined in
(\ref{measure4}) is just the integration over the $SL(2,R)$ group
manifold with the invariant measure,
\begin{eqnarray}
  \label{InvMeasure}
  d\mu(g) = \frac{1}{2\ 3!} \tr (g^{-1}dg)^3 = 2 d^4p\ \delta(1+l_P^2
  p^m p_m).
\end{eqnarray}
Hence we have
\begin{eqnarray}
  \label{Invariance}
  d\mu(g)=d\mu(gh)=d\mu(hg).
\end{eqnarray}
The above formulas will be used in computing one-loop diagrams in
section \ref{sec:1Loop}.

\section{Action and Feynman rules}
\label{sec:IntFeyn}

In this section, we construct the actions of interacting scalar
field theories in the noncommutative space-time and derive the 
Feynman rules\footnote{A similar derivation of action and Feynman rules
  was carried out for a deformed Minkowski space in \cite{luk}.}. 

A scalar field in the noncommutative space-time is defined by
associating momentum space wave functions $\Phi(g)$ to ``vertex
operators'' as follows:
\begin{eqnarray}
  \label{VertexOpeartor}
  \hat\Phi &=& \int d\mu(g) \Phi(g) e^{ik(g)\hat{x}}, \cr
  |\Phi\ra &=& \int d\mu(g) \Phi(g) e^{ik(g)\hat{x}}|0\ra
 = \int d\mu(g) \Phi(g) |p(g)\ra,
\end{eqnarray}
where $p(g)$ and $k(g)$ are defined by \eq{GroupElement}.
We impose the reality condition on the field $\hat\Phi$, 
i.e. $\hat\Phi^\dagger=\hat\Phi$.
We can define the product of the vertex operators by Hausdorff formula,
following the line of \cite{madore}.
Since the Hausdorff formula is nothing but the group multiplication, we obtain
\begin{eqnarray}
  e^{ik(g_1)\hat{x}}*e^{ik(g_2)\hat{x}} = e^{ik(g_1 g_2)\hat{x}}.
\end{eqnarray}
Making use of this $*$-product, we can construct interaction terms.
For example, the action of noncommutative $\phi^3$-theory can be
defined in the following way:
\be
  \label{action}
  S[\Phi] = \frac{-1}{2} \la 0 | \hat\Phi * (\hat{p}^2 + m^2)\hat\Phi
| 0 \ra+ \frac{\lambda}{3} \la 0|\hat\Phi *\hat\Phi *\hat\Phi |0 \ra,
\ee
where $\hat p^2=\hat p_\mu \hat p ^\mu$.

For practical computation, it is convenient to rewrite this action
in the momentum space representation with $SL(2,R)$ elements, and, for this,
we need to evaluate $\la 0 | e^{ik(g)\hat{x}} | 0 \ra=\la 0 | p(g) \ra$.
{}From the definition of the inner product, we can show that
$\la 0 | p(g) \ra$ is a $\delta$-function with respect to the
invariant measure $d\mu(g)$,
which has a support at the unit element of $SL(2,R)$:
\be
\la 0 | \Phi \ra \equiv \int d\mu(g) \la 0 | p(g) \ra \la p(g) | \Phi \ra
\Leftrightarrow \Phi(g=1) =\int d\mu(g) \la 0 | p(g) \ra \Phi(g).
\ee
Henceforth, we denote this $\delta$-function as $\delta(g)$.

Now we can write down the momentum representation of the action as
\begin{eqnarray}
  \label{ActionInMRep}
  S[\Phi] &=& \frac{-1}{2} \la 0 |\hat\Phi * (\hat{p}^2+ m^2)
  \hat\Phi | 0 \ra
+  \frac{\lambda}{3} \la 0 |\hat\Phi * \hat\Phi * \hat\Phi |0 \ra \cr
  &=& \frac{-1}{2} \int d\mu(g) \Phi(g^{-1}) (p^2(g) + m^2) \Phi(g) \cr
&& + \frac{\lambda}{3} \int \prod_{i=1}^3 d\mu(g_i)\delta(g_1 g_2
  g_3 ) \Phi(g_1) \Phi(g_2) \Phi(g_3) .
\end{eqnarray}
Assuming an appropriate path integral measure, we can perform
perturbative computation.
\begin{eqnarray}
  \label{PathIntegral}
  Z[J] &=& \int [D\Phi]\exp\left(iS_{\rm free}[\Phi] +iS_{\rm
  int}[\Phi]+i\int d\mu(g)J(g^{-1})\Phi(g) \right) \cr
  &=& \exp\left(i S_{\rm int}\left[ \frac{\delta}{i \delta J}\right]\right)
\exp \left( \frac{1}{2} \int d\mu(g)
\ i J(g^{-1})\frac{-i}{p^2(g)+m^2} i J(g) \right) \cr
&&  \times \int[D\Phi]\exp\left( \frac{i}{2} \int
  d\mu(g)(\Phi-\frac{1}{p^2+m^2}J)(p^2 + m^2)
(\Phi-\frac{1}{p^2+m^2}J) \right)
\end{eqnarray}
We may assume the last Gaussian path integral just gives a constant.
Thus, generalising to arbitrary $\phi^n$ interactions,
we obtain the following Feynman rules,
\begin{eqnarray}
  \label{FeynmanRule}
  {\rm propagator}: && \int d\mu(g)\frac{-i}{p^2(g)+m^2}, \cr
  n{\rm -vertex}    : && i \lambda_n \delta(g_1 \dots g_n)
\end{eqnarray}
Note that, since momentums are $SL(2,R)$ elements,
we have noncommutativity at the vertex.
In the following section, we compute some one-loop diagrams,
using these rules.

\section{One-loop computation}
\label{sec:1Loop}

In this section we shall compute the one-loop diagrams of the
two-point functions from $\phi^3$ and $\phi^4$-interactions defined
in the previous section.
We will show that the non-planar one-loop diagrams are finite and
have infrared singularities from the UV/IR mixing
\cite{minw}-\cite{raam}. We will also find that
those diagrams cause a problem concerning the conservation of momentums.
In this section, we set $l_P=1$ for simplicity.

\subsection{The planar diagram of the two-point function from $\phi^4$-interaction}
\label{sec: Planar}
\begin{figure}[htdp]
\begin{center}
\leavevmode
\epsfxsize=60mm
\epsfbox{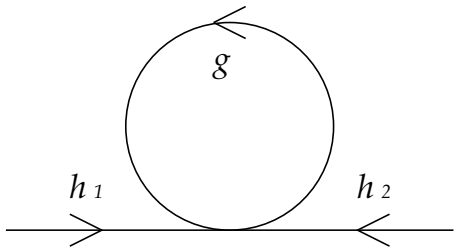}
\caption{The one-loop planar diagram of the two-point function.}
\label{fig:Planar}
\end{center}
\end{figure}

In the first place, we shall compute the simplest graph to show
our computational strategy:
the one-loop planar diagram of
the two-point function from $\phi^4$-interaction as in fig.(\ref{fig:Planar}).

{}From the Feynman rules in the preceding section,
the amplitude of this diagram is given by
\begin{eqnarray}
  \label{2PtPlanar1}
  \Gamma^{(2)}_P &=& \int d\mu(g) i\delta(g^{-1}h_2 h_1 g) \frac{-i}{p^2(g)
+m^2}
  = \delta(h_2 h_1) \int d\mu(g) \frac{1}{p^2(g)+m^2}.
\end{eqnarray}
The polar coordinate (\ref{PolarCoordinate}) is
convenient for the explicit evaluation of this integral,
and we further perform the change of the variable, $x=\ch^2 \chi$.
With this parameterisation, we obtain
\begin{eqnarray}
\label{PolarCoordinate2}
  p(g) &=& ( x^{\frac{1}{2}} \sin \theta,\ (1+x)^{\frac{1}{2}} \cos
  \phi ,\ (1+x)^{\frac{1}{2}} \sin \phi), \cr
  p^2(g) + m^2 &=& (p(g)_{-1})^2 -1+m^2 = x \cos^2 \theta - \cos^2 \mu,
  \cr
  d\mu(g) &=& \dg,
\end{eqnarray}
where $\sin\mu = m \ (0 \leq \mu \leq \pi/2 )$.

Since this integration turns out to be divergent, we introduce a
momentum cut-off $\Lambda$ in the $x$-integration.
\begin{eqnarray}
  \Gamma^{(2)}_P
&=& \delta(h_2 h_1) 
\int_0^{2\pi} d\phi \int_0 ^{2\pi} d\theta \int_1^{\Lambda^2} \frac{dx}{2} 
  \frac{1}{x\cos^2\theta-\cos^2\mu} \cr
&=& 2 \pi^2 \ \delta(h_2 h_1) 
 \int_1^{\Lambda^2} \frac{dx}{x} \int_0 ^{2\pi} 
\frac{i d\theta}{2 \pi i}
  \frac{1}{\cos^2\theta-\cos^2\mu_x} \cr
&=& 2 \pi^2 \ \delta(h_2 h_1) 
 \int_1^{\Lambda^2} \frac{dx}{x} \int_0 ^{2\pi} 
\frac{i d\theta}{2 \pi i}
  \frac{1}{\sin(\mu_x+\theta) \sin(\mu_x-\theta)},
\end{eqnarray}
where we have defined $\cos\mu_x = \cos\mu /x$.

\begin{figure}[t]
\begin{center}
\leavevmode
\epsfxsize=50mm
\epsfbox{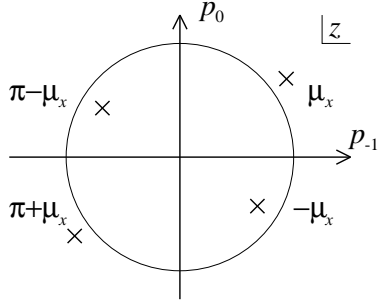}
\caption{$(m^2-i\epsilon)$-prescription}
\label{fig:pole}
\end{center}
\end{figure}

By the change of the integration variable $z=e^{i\theta}$,
the integration over $\theta$ 
is now a contour integration on a unit circle in the $z$-plane.
There are poles at $z =\pm e^{i\mu_x},\pm e^{-i\mu_x}$. 
Since they are on the unit circle, we adopt $(m^2-i\epsilon)$-prescription
to make the integral well-defined.
Then we just pick up the residues at $z = \pm e^{-i \mu_x}$
as shown in fig.(\ref{fig:pole}).
Thus the result of the integration is given by
\begin{eqnarray}
\label{2PtPlanar}
\Gamma^{(2)}_P
&=& 2 \pi^2 i \ \delta(h_2 h_1) 
 \int_1^{\Lambda^2} \frac{dx}{x} 
\frac{x}{\sqrt{x-(1-m^2)}\sqrt{1-m^2}} \cr
&=& - 4 \pi^2 i \ \delta(h_2 h_1) 
\left( \sqrt{\frac{m^2}{1-m^2}} -\sqrt{\frac{\Lambda^2}{1-m^2}}
\right) ,
\end{eqnarray}
where we have neglected the terms that vanish as $\Lambda \rightarrow \infty$.
We need to renormalize the divergence of the second term 
by a mass counter term.

\subsection{The non-planar diagram of the two-point function
from $\phi^4$-interaction}
\label{sec: NonPlanar}

\begin{figure}[tdp]
\begin{center}
\leavevmode
\epsfxsize=60mm
\epsfbox{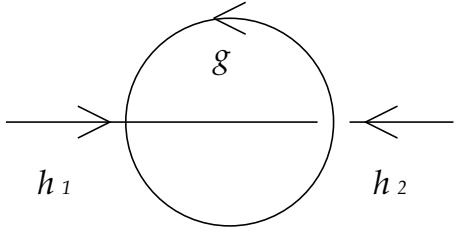}
\caption{The non-planar one-loop diagram of the two-point function}
\label{fig:NonPlanar}
\end{center}
\end{figure}

Next we consider the non-planar diagram of fig.(\ref{fig:NonPlanar}).
The amplitude is given by
\begin{eqnarray}
  \label{2PtNonPlanar}
  \Gamma^{(2)}_{NP} = \int d\mu(g) i\delta(h_2 g^{-1} h_1 g)
  \frac{-i}{p^2(g) +m^2}.
\end{eqnarray}
The integration over $g$ is performed for the solutions to
$h_2 g^{-1} h_1 g =1 $. Thus this integral is non-vanishing only
if $h_1$ and $h_2$ belong to the same conjugacy class, 
so both $p(h_1)_\mu$ and $p(h_2)_\mu$ must be
time-like or space-like, simultaneously.
We shall evaluate this integral when both the momentums are time-like.
Without loss of generality, we may assume $p(h_1)_\mu$ is in the time
direction, and $p(h_2)_\mu$ is represented by a vector obtained by
boosting a vector $p(h_2')_\mu$ which is in the time direction:
\begin{eqnarray}
  p(h'_2)_\mu &=& ( p(h'_2)_0,\ 0,\ 0), \cr
  p(h_2)_\mu &=& p(g_2^{-1}h'_2g_2) \cr
             &=& (p(h'_2)_0 \ \ch 2\chi_2,\ p(h'_2)_0
\ \sh 2\chi_2 \sin\phi_2,\ -p(h'_2)_0 \ \sh 2\chi_2 \cos\phi_2).
\end{eqnarray}
We can rewrite
\begin{eqnarray}
\Gamma^{(2)}_{NP} &=& \int d\mu(g) \delta( h'_2 g_2 g^{-1} h_1 g g_2^{-1})
\frac{1}{p^2(g) + m^2} \cr
&=& \int d\mu(g) \delta(h'_2 g^{-1} h_1 g)
\frac{1}{p^2(gg_2)+m^2},
\end{eqnarray}
where we have changed the integration variable $g$ to $g g_2$.
An adjoint action is simplified when acting on a vector in the
time-direction:
\begin{eqnarray}
  p(g^{-1}h_1g)_0 &=& p(h_1)_0 (1+2(p_1^2+p_2^2)), \cr
  p(g^{-1}h_1g)_i &=& 2 p(h_1)_0 (p_0 p_i +\epsilon_i{}^j p_{-1} p_j),
\end{eqnarray}
where we have abbreviated $p_m = p(g)_m$.
We can show the following formula
\footnote{Abbreviating $p(g)=p$ and $\ p(h)=q$,
\begin{eqnarray}
\int d\mu(g) \delta(gh) \phi(g)
&=&  2 \int d^4 p \ \delta(1 + p^m p_m) \ |p_{-1}|
\theta(p_{-1} q_{-1}) \ \delta^{(3)}(p+q) \phi(p) \cr
&=& 2 \int dp_{-1} \ \delta(q_{-1}^2 -p_{-1}^2)\ |p_{-1}|
 \theta(p_{-1} q_{-1}) \ \phi(p_{-1},-q_{\mu}) \cr
&=& \int dp_{-1} \{ \delta(q_{-1} - p_{-1})+\delta(q_{-1} + p_{-1}) \}
 \theta(p_{-1} q_{-1}) \ \phi(p_{-1},-q_{\mu}) \cr
&=& \phi(q_{-1},-q_\mu) = \phi(h^{-1}),
\nonumber
  \end{eqnarray}
where the step function $\theta(p_{-1} q_{-1})$ is needed
to discard the contribution from $g =-h$.}:
\begin{eqnarray}
\delta(gh) = |p(h)_{-1}| \
\theta(p(g)_{-1} p(h)_{-1})\ \delta^{(3)}(p(g) + p(h)).
\end{eqnarray}
Making use of it, we obtain
\begin{eqnarray}
\Gamma_{NP}^2
&=& | p(h_1)_{-1}| \ 2 \int d^4p \ \delta(1+p^m p_m)\cr
&& \ \times \theta(p(g^{-1} h_1 g)_{-1} p(h'_2)_{-1})
\  \delta^{(3)}(p(g^{-1} h_1 g) + p(h'_2)) \frac{1}{p^2(g g_2) +m^2} \cr
&=& |p(h_1)_{-1}| \ 2 \int d^4p \ \delta(1+p^m p_m) \cr
&& \ \times \theta(p( h_1 )_{-1} p(h'_2)_{-1})
\ \delta(p(h_1)_0 (1+ 2 p_1^2 + 2 p_2^2)+p(h'_2)_0) \cr
&& \ \times \delta(2p(h_1)_0 (p_0p_1-p_{-1}p_2))
\ \delta(2p(h_1)_0 (p_0p_2+p_{-1}p_1)) \frac{1}{p^2(g g_2) + m^2}.
\end{eqnarray}
The integrations over $p_1$ and $p_2$  can be trivially performed by
the 2nd and 3rd $\delta$-functions, picking up the values $p_1=p_2=0$
and giving the Jacobian factor
$1/4(p(h_1)_0)^2 (p_{-1}^2+ p_0^2)=1/(-4p^2(h_1))(p_{-1}^2+p_0^2)$:
\begin{eqnarray}
\Gamma^{(2)}_{NP}
&=& \frac{|p(h_1)_{-1}|}{- 4 p^2(h_1)}
\  \theta(p(h_1)_{-1} p(h'_2)_{-1}) \ \delta(p(h_1)_0 + p(h'_2)_0) \cr
&&  \ \times 2 \int dp_{-1}dp_0 \ \delta(1-p_{-1}^2-p_0^2)
\left. \frac{1}{p^2(g g_2) + m^2}
\right|_{p_1=p_2=0}.
\end{eqnarray}
Thus the integration reduces to the contour integral on a
unit circle in $(p_{-1},p_0)$-plane.
We evaluate this integral by $(m^2-i\epsilon)$-prescription:
\begin{eqnarray}
  \Gamma^{(2)}_{NP}
&=&  \frac{|p(h_1)_{-1}|}{- 4 p^2(h_1)}
\  \theta(p(h_1)_{-1} p(h'_2)_{-1})\ \delta(p(h_1)_0 +
p(h'_2)_0) 
\int_0^{2\pi}d\theta \frac{1}{\ch^2\chi_2
  \cos^2(\theta+\frac{\phi_2}{2}) + \cos^2\mu} \cr
&=& \frac{|p(h_1)_{-1}|}{- 4 p^2(h_1)}
\ \theta(p(h_1)_{-1} p(h'_2)_{-1})\ \delta(p(h_1)_0 + p(h'_2)_0)
 \frac{2\pi i}{\sqrt{(\sh^2\chi_2+\sin^2\mu)\cos^2\mu}}
\end{eqnarray}

Taking into account the relation
$\sh^2\chi_2=-\frac{1}{2}\left(\frac{p(h_1)\cdot p(h_2)}{p(h_1)^2}+1\right)$
and an identity,
\begin{eqnarray}
 \theta(p_{-1} q_{-1})\ \delta(p_0 + q_0)
= \frac{\sqrt{-p^2}}{|p_{-1}|}
\theta(-p_0 q_0) \ \delta(p_{-1}-q_{-1}), \cr
\end{eqnarray}
we finally obtain the following Lorentz invariant result:
\be
i \Gamma^{(2)}_{NP} = \frac{- \pi}{\sqrt{2(1-m^2)}}
\frac{\theta(-p(h_1)_0 p(h_2)_0) \ \delta (p(h_1)_{-1}-p(h_2)_{-1})}
{\sqrt{(1-2m^2)p^2(h_1)+p(h_1)\cdot p(h_2)}}.
\ee

This result shows that the non-planar diagram has an infrared
singularity coming from the UV/IR mixing \cite{minw}-\cite{raam}
irrespective of massive theory.
We can also see that the momentum is not conserved while the 
momentum square is.
One might expect that the momentum would be conserved in the
commutative limit $l_p \rightarrow 0$, since, in this limit,
the commutation relations
\eq{NCPhaseSpace} have the translational symmetry $\hat x^\mu\rightarrow
\hat x^\mu+v^\mu$ with a $c$-number vector $v^\mu$. However this is not true.
In the commutative limit, $m$ approaches to zero, and the
factor $1/\sqrt{p^2(h_1)+p(h_1)\cdot p(h_2)}$ in the amplitude
$\Gamma^{(2)}_{NP}$ has an infinite peak at $p_\mu(h_1)+p_\mu(h_2)=0$. 
Although the momentums are conserved at
this peak, we can easily show that the factor $1/\sqrt{p^2(h_1)+p(h_1)\cdot
p(h_2)}$ is not a $\delta$-function under the measure $d\mu(g)$.
Thus, in this theory, the momentum conservation is violated
substantially even in the low-energy limit. The problem of the 
violation of momentum conservation was observed in another
noncommutative space-time in \cite{luk}. 
We shall discuss this problem further in section \ref{sec:trans}.

\subsection{The two-point function from $\phi^3$-interaction}

\begin{figure}[htdp]
\begin{center}
\leavevmode
\epsfxsize=60mm
\epsfbox{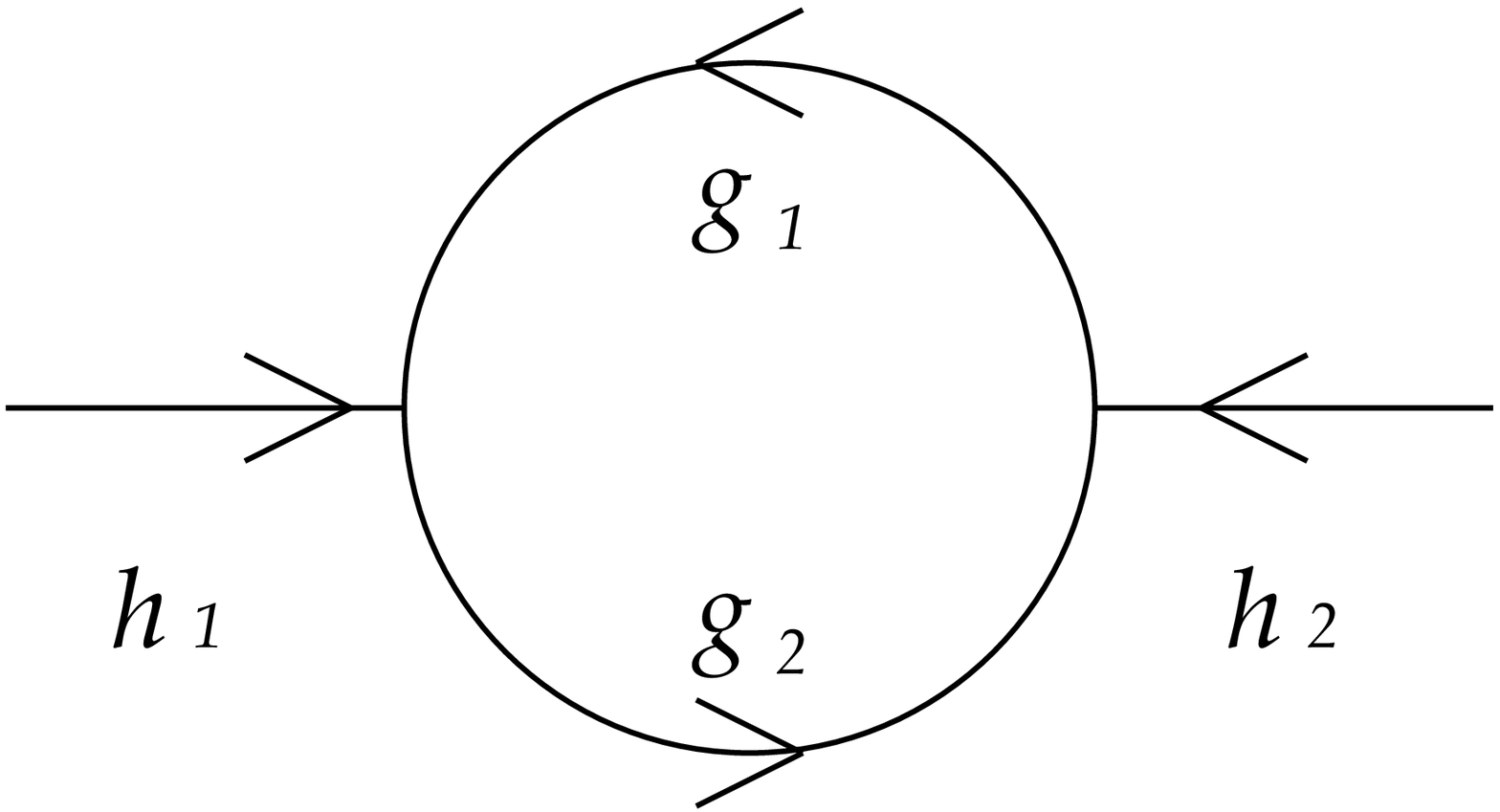}
\hspace{4em}
\epsfxsize=60mm
\epsfbox{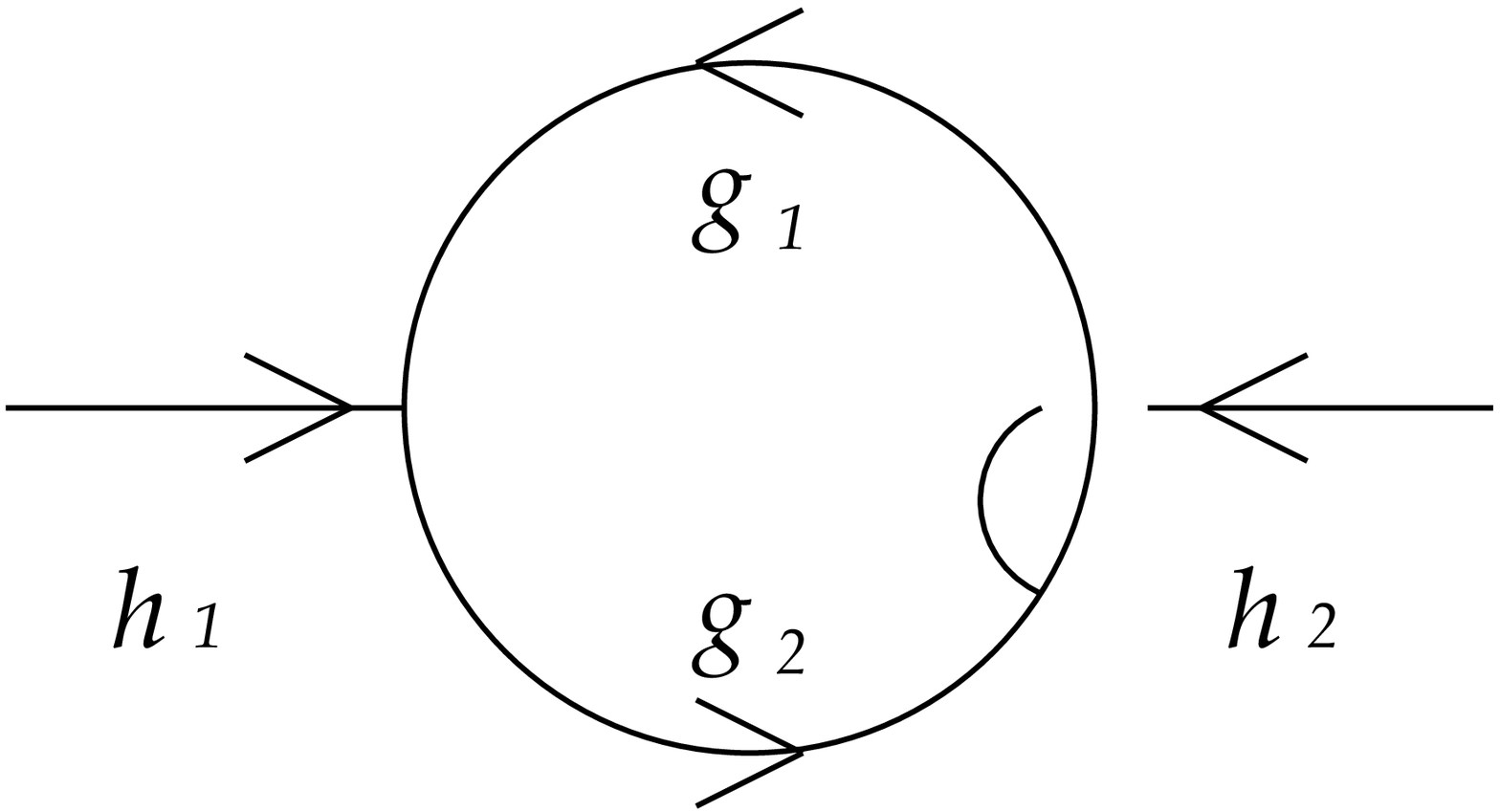}
\caption{(a) The one-loop planar diagram \hspace{3em}
(b) The one-loop non-planar diagram }
\label{fig:DiagramPhi3}
\end{center}
\end{figure}

The one-loop contributions to the two-point function from $\phi^3$-interaction
can be computed in the same way as in $\phi^4$-interaction.
There are two diagrams as in figs.(\ref{fig:DiagramPhi3}).

The contribution from the diagram (\ref{fig:DiagramPhi3})-(a) is
given by
\begin{eqnarray}
\label{eq:gammtwo}
\Gamma^{(2)}_P
&=&  \int d\mu(g_1) d\mu(g_2) i\delta(g_2^{-1} h_1 g_1)
i \delta(g_1^{-1} h_2 g_2)
\frac{-i}{p^2(g_1)+m^2} \frac{-i}{p^2(g_2)+m^2}   \cr
&=&  \int d\mu(g) \delta(h_2 h_1)
\frac{1}{p^2(g)+m^2} \frac{1}{p^2(h_1 g)+m^2}.
\end{eqnarray}
In this case the three-momentum $p_\mu$ conserves, so we may assume,
without loss of generality,
$p(h_1)_\mu = - p(h_2)_\mu = ( \sin \theta_1,\ 0,\ 0)$ for time-like
three-momenta.
Parameterising the loop momentum by the polar coordinate as in
eq.(\ref{PolarCoordinate2}), we have
$p^2(h_1 g) = \ch^2 \chi \cos^2 (\theta + \theta_1) -1$.
Thus
\begin{eqnarray}
\label{NonPlanar3-1}
\Gamma^{(2)}_P
&=&  \delta(h_2 h_1) \dg \frac{1}{x \cos^2 \theta - \cos^2 \mu}
\frac{1}{x \cos^2(\theta +\theta_1)-\cos^2 \mu} \cr
&=& \frac{\pi }{1-m^2} \delta(h_2 h_1)
\frac{-1}{\sin \theta_1} I(\theta_1),
\bee
where
\bea
\label{eq:defofi}
I(\theta_1) = \int_0^{2\pi} d\theta
\frac{1}{\sin2\theta}
\log \left( \frac{\cos^2(\theta -\frac{1}{2} \theta_1)
\sin(\theta+\frac{1}{2}\theta_1+\mu)
\sin(\theta+\frac{1}{2}\theta_1-\mu)}
{\cos^2(\theta + \frac{1}{2} \theta_1)
\sin(\theta-\frac{1}{2}\theta_1+\mu)
\sin(\theta-\frac{1}{2}\theta_1-\mu)}
\right).
\end{eqnarray}

\begin{figure}[htdp]
\begin{center}
\leavevmode
\epsfxsize=60mm
\epsfbox{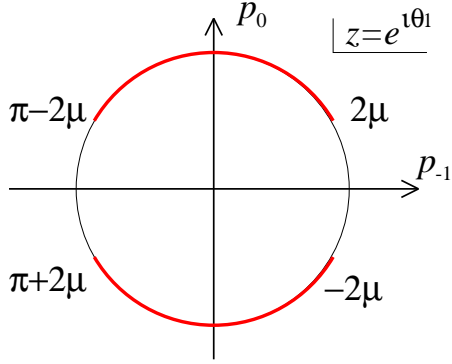}
\caption{The bold line corresponds to the continuous spectrum of 
two-body states}
\label{fig:Cut}
\end{center}
\end{figure}

Now let us compute the first derivative $I'(\theta_1)$ with respect to 
$\theta_1$.
The $\theta$-integral can be rewritten as a contour
integral on a unit circle in the $z=e^{i\theta}$ plane.
There appears several poles on the contour.
These poles originate from the poles of the propagators in
\eq{NonPlanar3-1} at $\cos(\theta \pm \frac12 \theta_1)=\pm
\sqrt{(1-m^2)/x}$, where we have shifted $\theta$ by $-\frac12 \theta_1$
from that of \eq{NonPlanar3-1}.
These poles are treated by the $(m^2-i\epsilon)$-prescription as
the preceding subsections.
Then, since the poles of $I'(\theta_1)$ at
$\cos(\theta \pm \frac{1}{2} \theta_1)=0$ come from the
$x=\infty$ contributions of the integral \eq{NonPlanar3-1}, we see that
these poles are in fact degenerate pairs of poles in the inside and those in
the outside of the unit circle.
By an explicit computation, it turns out that the contributions from those
poles cancel.
Out of the poles of $I'(\theta_1)$ at 
$\sin(\theta\pm\frac12\theta_1\pm\mu)=0$,
which come from the $x=1$ contributions of the integral
\eq{NonPlanar3-1}, only the poles at
$\sin(\theta\pm\frac12\theta_1+\mu)$ are inside.
Evaluating the residues, we obtain
\begin{eqnarray}
  I'(\theta_1)
&=& 2 \pi i \left(
\frac{1}{\sin(\theta_1-2\mu)} - \frac{1}{\sin(\theta_1+2\mu)}
\right).
\end{eqnarray}
Using $1/\sin x = (\log\tan(x/2))'$, we obtain
\begin{eqnarray}
 I(\theta_1)
&=& 2 \pi i \log \left( \frac{\tan(\mu - \frac{1}{2} \theta_1)}
{\tan(\mu + \frac{1}{2} \theta_1)}\right),
\end{eqnarray}
where the integration constant is determined by $I(0)=0$.
Thus we reach the result
\begin{eqnarray}
\label{NonPlarnar3-Result}
i \Gamma^{(2)}_P &=& \frac{2\pi^2}{1-m^2} \delta(h_2h_1)
\frac{1}{\sin\theta_1} \log \left( \frac{\tan(\mu-\frac{1}{2}
    \theta_1)}
{\tan(\mu+\frac{1}{2} \theta_1)} \right),
\end{eqnarray}
where $\mu $ should be understood as $\mu-i\epsilon$.
This expression has the branch cuts which represent the continuous
spectrum of two-body states as in the fig.(\ref{fig:Cut}).

As in $\phi^4$-interaction, the contribution from
fig.(\ref{fig:DiagramPhi3})-(b) also violates the conservation of the
three-momentum. We have computed the amplitude in the case that
the external momenta are time-like.
The result is
  \begin{eqnarray}
    i\Gamma^{(2)}_{NP}
&=& \frac{\pi }{\sqrt{1-m^2}}
 \theta(-p_0 p'_0) \
\delta(p_{-1} - p'_{-1})  \cr
&& \times \frac{1}{\sqrt{\frac{1}{4}(p+p')^2-m^2p^2}}
\ \frac{2(1-m^2)p^2 - \frac{1}{4}(p+p')^2 }
{-(\frac{1}{4} (p+p')^2)^2 - 4 (1-m^2)(\frac{1}{4}(p+p')^2 - m^2 p^2)}
  \end{eqnarray}
where $p_m=p(h_1)_m,\ p'_m=p(h_2)_m$.
This amplitude has the same problems discussed for the non-planar
diagram from $\phi^4$-interaction.

\subsection{Some comments on higher loops and renormalization}

In the preceding subsections, we see that the one-loop contributions
to the two-point function conserve $p^2$ while they do not conserve $p_\mu$.
In higher loops, the $p^2$ is not conserved either.
To see this, let us consider the diagram of fig.(\ref{fig:p2violation})
as an example.
The amplitude is proportional to
\be
\int d\mu(g_1) d\mu(g_2) d\mu(g_3) \delta(g_1 g_2^{-1}h_2h_1)
  \delta(g_1^{-1} g_3 g_2 g_3) 
\sim \int d\mu(g_1) d\mu(g_3) \delta(g_1 g_3^{-1} g_1^{-1} g_3 h_2
  h_1).
\ee
Thus $h_1$ and $h_2$ do not need to belong to the same conjugacy class, 
and therefore $p^2$ is not conserved.

\begin{minipage}[c]{0.35\textwidth}
\leavevmode
\epsfxsize=50mm
\epsfbox{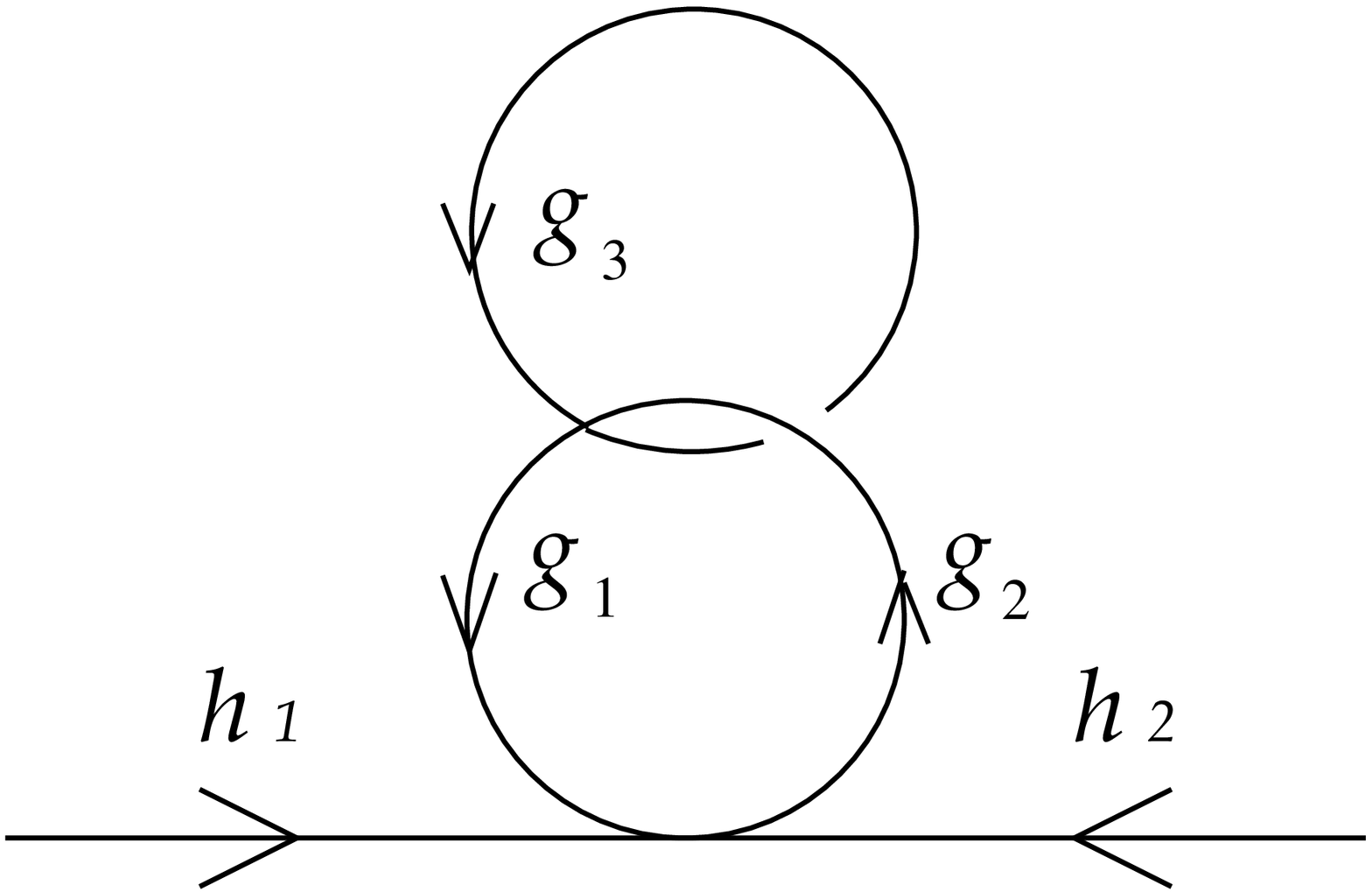}\\
\refstepcounter{figure}
Figure \thefigure: A $p^2$-violating graph
\addcontentsline{lof}{figure}{\protect\numberline{\thefigure}
{A $p^2$-violating graph}}
\label{fig:p2violation}
\end{minipage}
\begin{minipage}[c]{0.3\textwidth}
\leavevmode
\epsfxsize=38mm
\epsfbox{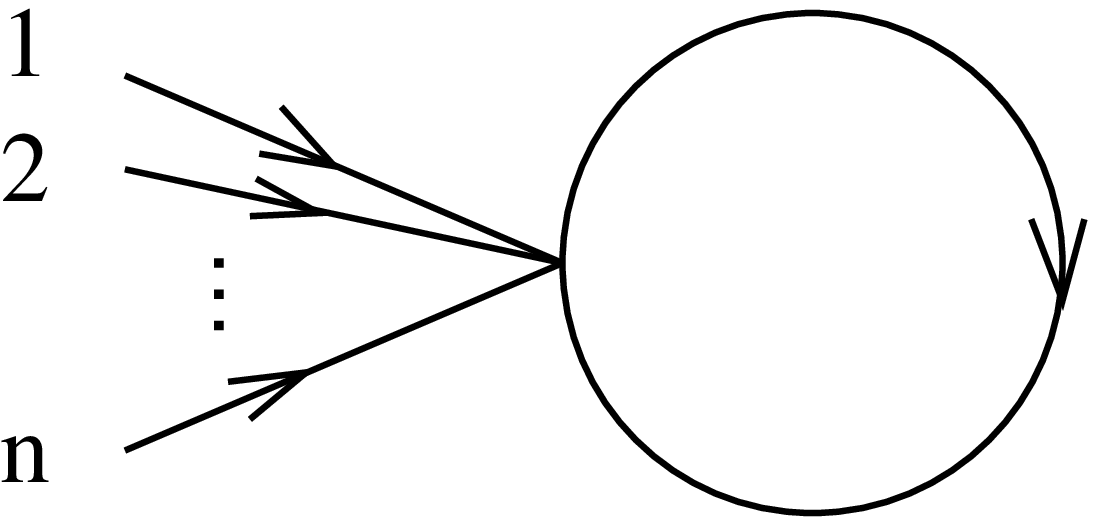}\\
\refstepcounter{figure}
Figure \thefigure : A divergent tadpole graph
\addcontentsline{lof}{figure}{\protect\numberline{\thefigure}{divergent tadpole graph}}
\label{fig:tad}
\end{minipage}
\begin{minipage}[c]{0.3\textwidth}
\hspace{20mm}
\epsfxsize=45mm
\epsfbox{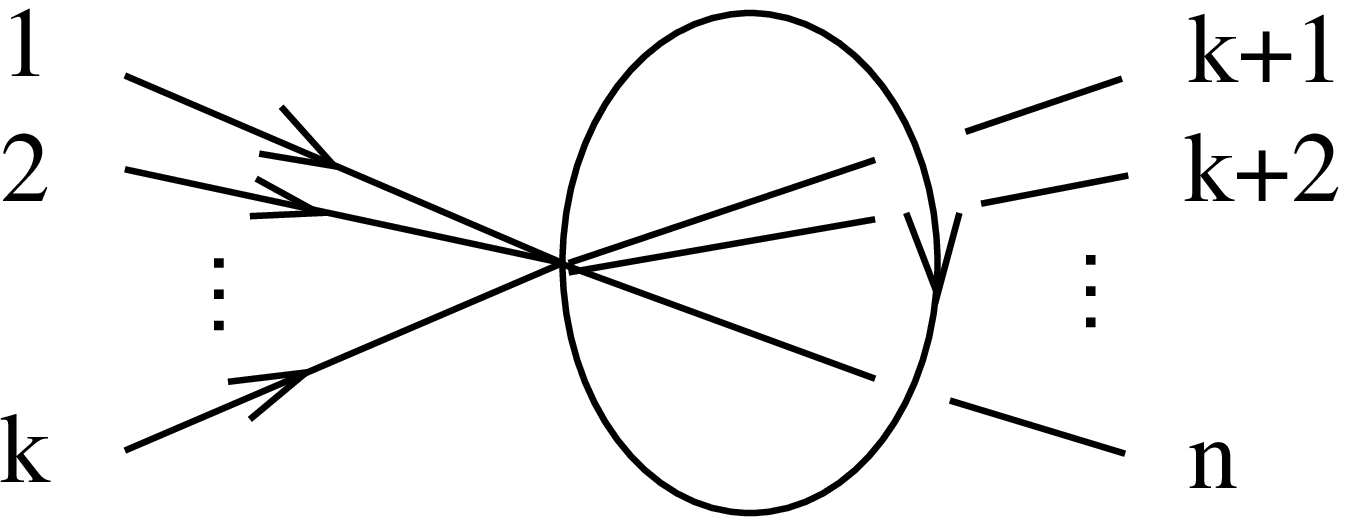}\\
\refstepcounter{figure}
Figure \thefigure : A non-planar tadpole graph
\addcontentsline{lof}{figure}{\protect\numberline{\thefigure}{divergent tadpole graph}}
\label{fig:nptad}
\end{minipage}

As for the renormalization at one-loop order,
the only divergent graphs 
are planar tadpole graphs such as fig.(\ref{fig:tad}).
These can be calculated simply by replacing 
the external momenta of the result of subsection \ref{sec: Planar}.
Then their external momentum dependence becomes $\delta(h_1 \dots h_n)$, 
and these divergences can be absorbed by $n$-point counter terms and
renormalizable. 
No non-planar graphs such as fig.(\ref{fig:nptad}) diverge. 
This is because those graphs can be evaluated 
only by replacing the external 
momenta  $(h_1,h_2)$ of the result of subsection \ref{sec: NonPlanar}
with those of appropriate channel $(h_1 \dots h_k,h_{k+1}\dots h_n)$.
Notice that the group manifold $\delta$-function at the vertex reduces the
dimensions of loop-momentum integration 
and has an effect similar to the Moyal phase
factor which serves as a damping factor. 
For example, in the non-planar graphs such as figs.(\ref{fig:NonPlanar}) 
and (\ref{fig:nptad}), the loop momentum runs only over the one-dimensional
subgroup of $SL(2,R)$ determined from the external momenta.

Beyond the one-loop order, we do not know whether 
the noncommutative scalar field theory we are discussing is 
renormalizable or not.
In the case of scalar field theory with the Moyal type noncommutativity,
renormalizability is shown up to two-loop \cite{rensca} and also 
a convergence theorem of Feynman integral has recently been well 
understood \cite{ChepRoi}.
For the noncommutative YM and QED cases, renormalizability is shown 
up to one-loop \cite{renone}-\cite{renthree}, \cite{haya}.  
In order to investigate the issue in our field theory from perturbative
analysis,  we need first a systematic evaluation of Feynman integral as was
done in \cite{ChepRoi}.
In the present work, this is out of our scope.

\noindent
\section{Translational symmetry}
\label{sec:trans}

In the previous section, we have shown that the contributions from the
non-planar diagram violates the momentum conservation.
This violation comes from the fact that the commutation relations of
the coordinate operators \eq{NCPhaseSpace}
do not respect the naive translational
invariance, $\hat{x}^\mu \rightarrow \hat{x}^\mu+v^\mu$ with a c-number vector
$v^\mu$.
Since the translational invariance is recovered in the commutation relations
in the commutative limit $l_P \rightarrow 0$, one naively expects that
the violation should vanish in this limit. But this naive expectation
is not true as we saw for the non-planar one-loop
results in the preceding section.
Since this violation exists substantially in the low energy limit
while the momentum conservation is one of the fundamental
components in the present theoretical physics, we need an exact
symmetry in the noncommutative space-time, which corresponds to the
translational symmetry in the commutative limit.

To introduce such symmetry, let us define a field
\be
\hat{\Psi}=\sum_{n=0} \int d\mu (g) \ \Phi (g)_{\mu_1\cdots\mu_n}\
\hat{p}^{\mu_1}\cdots \hat{p}^{\mu_n} \exp(i k_\mu(g) \hat{x}^\mu),
\ee
where $\Phi(g) _{\mu_1\cdots\mu_n}$ are c-number tensor fields.
This can be regarded as a non-local generalisation of the
scalar field defined in \eq{VertexOpeartor}.

Let us define a unitary operator
\be
U(v)=\exp(iv_\mu \hat{p}^\mu).
\ee
The $p_\mu=0$ eigen-state $|0\rangle$ is invariant under this unitary
operator
\be
U|0\rangle=|0\rangle.
\ee
We define the translational transformation of the field $\hat \Psi$ by
\be
\hat{\Psi}\rightarrow U^\dagger \hat{\Psi} U.
\ee
Then this translational transformation is a symmetry of the following
type of action
\be
S=\sum \langle 0 | (\hat{\Psi}^\dagger)^m f(\hat{p}) \hat{\Psi}^n |0\rangle.
\label{eq:actiontensor}
\ee

In usual commutative cases, the translational transformation just
generates a phase shift $\exp(-ik_\mu v^\mu)$, and the tensor fields
$\hat \Phi (g)_{\mu_1\cdots\mu_n}$ do not mix. 
Hence we may truncate the fields only to the scalar sector. 
However, in our noncommutative case, 
since the commutation relations are given by \eq{NCPhaseSpace}, the
tensor fields mix. Thus the translational symmetry transforms local
fields into non-local ones in general.

This coordinate dependence of locality would not be ridiculous
when we recall the initial motivation of considering the
noncommutative space-time. One of the authors regarded the
noncommutative space-time \eq{NCPhaseSpace} as a space-time realizing 
a space-time uncertainty relation derived from a gedanken 
experiment \cite{sasa2}. 
We have no means to obtain a coordinate system more accurate than
the limit specified by the space-time uncertainty relation.
The uncertainty will become larger if a space-time event is farther from
the origin of the coordinate system. Thus the space-time spread of
an event is not an invariant notion anymore under the translational
symmetry.

In places where the general relativity and quantum mechanics play
major roles simultaneously,
similar kinds of issues have already been observed in the literatures.
The subtlety of the locality of an event has appeared already as an
aspect of the black hole complementarity of Susskind \cite{revsuss}.
For instance, 
he argued that the location of the baryon violation in a black hole
geometry is an observer dependent notion.
Moreover, in the holographic description of the world \cite{holo,holosuss},
a space-time event is mapped to a screen, and the spread of the image
will depend on the relative locations of the event and the
screen \cite{revsuss}.

Lastly we see that the action \eq{eq:actiontensor} may be derivable
from a pregeometric 
action\footnote{A pregeometric action in string theory was proposed 
in \cite{pre}.} in the form
\be
S_{preg}=\langle 0 | g(\hat{\Psi}^\dagger,\hat{\Psi}) |0\rangle.
\label{eq:preaction}
\ee
An action with a kinetic term would be generated in a certain
background of the field $\hat{\Psi}=\hat{\Psi}_0$.
For example, starting from a cubic pregeometric action and considering
the background of $\hat{\Psi}_0$ with $\Phi_0=m^2$ and 
$\Phi_0^{\mu\nu}=\eta^{\mu\nu}$, we obtain an action with a kinetic
term and cubic interactions for the fluctuation field $\hat{\Psi}_1$ in
$\hat \Psi=\hat \Psi_0+ \hat \Psi_1$.
The action \eq{eq:preaction} would be more interesting
than \eq{eq:actiontensor}, since it has more symmetry.

\section{Thermodynamics}
\label{sec:thermo}

In this section, we will discuss an analogue of thermodynamics of 
free scalar field theory in the noncommutative space-time.
Reflecting the noncommutativity of the space-time, 
we find a non-trivial behaviour at high temperature.
Following the common trick in usual commutative field theories,
we will take the Euclidean metric and calculate the partition
function, 
imposing the periodicity of the inverse temperature in the time-direction.

In the Euclidean metric with signature $\eta^{\mu\nu}=(1,1,1)$
\footnote{The results in this section are the same for the signature 
$\eta^{\mu\nu}=(-1,-1,-1)$.},
the consistency with the Jacobi identity changes 
the last commutation relation of \eq{NCPhaseSpace} to
\be
[\hat{p}^{\mu},\hat{x}^{\nu} ] =
-i \eta^{\mu\nu} \sqrt{ 1 - l_P^2 \hat{p}^2 } +
  i l_P  \epsilon^{\mu\nu\rho} \hat{p}_\rho.
\ee
Thus the coordinate operators in the momentum representation are given by
\be
\hat{x}^\mu=i\sqrt{1-l_P^2p^2}\frac{\pt}{\pt p_\mu}
-i l_P \epsilon^{\mu\nu\rho}p_\nu\frac{\pt}{\pt p^\rho},
\ee
where we have ignored the possible ambiguity \cite{sasa2}.

We now use a parameterisation
\bea
p_0&=&\frac1{l_P} \cos \omega \sin \theta, \cr
p_1&=&\frac1{l_P} \sin \omega \cos \varphi, \cr
p_2&=&\frac1{l_P} \sin \omega \sin \varphi,
\bee
where the ranges of the parameters are given by
\be
0\leq \theta <2\pi,\ 0\leq \omega \leq \frac\pi{2},\ 0\leq \varphi < 2\pi.
\ee
Then the measure is given by
\be
(1-\lp^2 p^2)^{-\frac12}d^3p=d\theta d\omega d\varphi \sin\omega \cos\omega.
\ee
Using this parameterisation, the coordinate operators are rewritten as
\bea
x^0&=& il_P \left( \frac\pt{\pt\theta}-\frac\pt{\pt
    \varphi}\right), \cr
x^1&=& il_P \left( \tan \omega \sin(\theta-\varphi)
  \frac\pt{\pt\theta}+\cos(\theta-\varphi)\frac\pt{\pt \omega}
-\cot \omega \sin (\theta-\varphi)\frac\pt{\pt \varphi} \right), \cr
x^2&=& il_P \left( \tan \omega \cos(\theta-\varphi)
  \frac\pt{\pt\theta}+\sin(\varphi-\theta)\frac\pt{\pt \omega}
+\cot\omega \cos (\theta-\varphi)\frac\pt{\pt \varphi} \right).
\label{eq:operat}
\bee

Because of the noncommutativity of the coordinate operators, it is
impossible to find an operator which shifts $x^0$ by a certain
c-number while $x^{1,2}$ remain intact.
Below we shall just find an operator which shifts $x^0$ by $\beta$ but
does not change the expectation values 
$\langle \phi | x^{1,2} | \phi \rangle$, where $|\phi\rangle$
is an eigen-state of $x^0$.

{}From \eq{eq:operat}, we see that the eigen-function of the operator
$x^0$ is given by
\be
\phi=e^{id(\theta-\varphi)} f(\theta+\varphi,\omega).
\ee
The periodicities of the wave function under
$\theta\rightarrow \theta+2\pi,\ \varphi\rightarrow
\varphi+2\pi$ constrain the values of $d$ as
\be
d=\frac{n_d}2,
\ee
where $n_d$ is an integer, and leads to that $f$ must satisfy
\be
f(y,\omega)=(-1)^{n_d} f(y+2\pi,\omega).
\ee

A candidate of the shift operator with $U^\dagger x^0 U = x^0+\beta$ 
is given by
\be
U(\beta)=\exp\left(-\frac {i}{l_P}\beta\theta \right),
\label{eq:shift}
\ee
where, in order to satisfy the periodicity with respect to $\theta$,
\be
\beta=l_P n_\beta
\label{eq:quantbeta}
\ee
with a positive integer $n_\beta$.
Since $U^\dagger x^1 U = x^1 +\beta \tan\omega \sin(\theta-\varphi)$,
we find
\bea
\langle \phi | \delta x^1 |\phi\rangle &=&
\langle \phi | (U^\dagger x^1 U -x^1)|\phi\rangle \cr
&=& \beta \int d\theta d\omega d\varphi \sin\omega \cos\omega
|f(\theta+\varphi,\omega)|^2
\tan\omega \sin(\theta-\varphi)\cr
&=&0.
\bee
Similarly $\langle \phi | \delta x^2 |\phi\rangle=0$.
The shift operator $U$ could contain $\varphi$-derivative in order to 
satisfy only the above features. 
However, since $\varphi$-derivative generates a spatial rotation
which we do not need,
we assume that \eq{eq:shift} is the appropriate operator.

The thermodynamic partition function for the inverse temperature
$\beta$ is obtained by taking the trace with respect to the
states satisfying $U|\phi\rangle=|\phi\rangle$. Hence we 
sum up over the states with $\theta=2\pi\lp n /\beta$, 
where $n$ is an arbitrary integer. Thus we obtain
\bea
\ln Z&=&-\frac12 {\rm Tr}_{U=1}\left(\ln(p^2+m^2)\right) \cr
&=& - \frac1{8\pi} \sum_{n=-\infty}^\infty \int d\theta d^2q\
\delta\left(\theta-\frac{2\pi\lp n}{\beta}\right)
\ln\left(\frac1{\lp^2}\sin^2\theta + q^2\cos^2\theta +m^2 \right).
\label{eq:part}
\bee
The overall numerical constant is determined by comparing with the usual
commutative case in the limit
$\beta\rightarrow\infty$\footnote{Since the momentum space is doubly
degenerate, we equate \eq{eq:part}
in the $\beta\rightarrow\infty$ limit with the usual
thermodynamic partition function of two real scalar fields.}.
The integration over $\theta$ can be done by the contour
integration with the new variable $z=e^{i\theta}$ after applying
the formula $\sum_m e^{2\pi mti}=\sum_n \delta(t-n)$ to \eq{eq:part}.
The result is
\bea
\ln Z&=&-\frac1{2\pi}\int_0^{\frac1{\lp}} qdq \ln
\left( y_+^{\beta/2\lp}-y_+^{-\beta/2\lp}\right), \cr
y_+&=&\sqrt{1+\frac{-q^2+1/\lp^2}{q^2+m^2}}+
\sqrt{\frac{-q^2+1/\lp^2}{q^2+m^2}}.
\label{eq:partq}
\bee

The logarithm of the partition function \eq{eq:partq} has the usual 
behaviour $\sim -1/\beta^2$ in the low temperature. If we ignore the
quantisation of $\beta$ \eq{eq:quantbeta}, it has the behaviour
$\sim -\log\beta$ at the high temperature $\beta \ll \lp$.
This suggests the reduction of the degrees of freedom in
the high energy region.
If we can apply the usual first law of thermodynamics to this system,
we obtain the entropy density $s\sim -\ln\beta$ and the energy
density $e\sim 1/\beta$ at the high temperature $\beta \ll \lp$.
Thus $s/\sqrt{e}$ decreases when the temperature increases.
On the other hand, in the low temperature region,
$s/\sqrt{e}\sim 1/\sqrt{\beta}$, and it increases with the
temperature. Thus there is an upper bound of $s/\sqrt{e}$, in
agreement with the qualitative argument given in \cite{sasa1}.

However, it is not clear whether we can really regard $\beta$ as the inverse
of the temperature, and can use the first law of thermodynamics to
obtain the energy and entropy. To have a reliable discussion on
these issues, we have to construct statistical thermodynamics in the
noncommutative space-time.
To do so, we need a Hamiltonian and its spectra.
Moreover, in usual commutative cases, a thermodynamic system is put in an
imaginary box
and the states are well regularized by the infrared cut-off given by
the size of the box.
Since these issues are non-trivial in the noncommutative
space-time, the construction of statistical thermodynamics 
remains unsolved.

\section{Summary and discussions}
\label{sec:sum}

In this paper, we have analysed scalar field theories in the noncommutative
three-dimensional space-time characterised by 
$[x^\mu,x^\nu]=2i\lp\epsilon^{\mu\nu\rho}x_\rho$.
The one-particle Hilbert space in momentum representation
is represented by the $SL(2,R)$ group manifold.
Since momentums are elements of $SL(2,R)$, the interaction vertices
have definite ordering of legs coming from the noncommutativity of the group
elements.
We have performed some one-loop computations, which can be evaluated
explicitly by contour integrations in complex planes.
The non-planar one-loop diagrams of the two-point functions from
$\phi^4$ and $\phi^3$-interactions were 
shown to be finite and have the infrared
singularity coming from the UV/IR mixing.

The most peculiar feature of the noncommutative space-time
is that the commutation relations among the coordinates
do not respect the translational symmetry. This violation of
translational symmetry is natural from the view point of
a fuzzy space-time \cite{sasa2}. If an event is farther from the
origin of a reference frame, the location of the event will become
more fuzzy by the quantum fluctuation of the space-time. Since the
translational symmetry is recovered in the $\lp\rightarrow 0$ limit
in the commutation relations \eq{NCPhaseSpace},
one might expect that the momentum conservation would be recovered in this
limit. However we have shown explicitly that this is not true for
the non-planar contributions.
To remedy this defect, we had to introduce an infinite
number of tensor fields.
We leave the analysis of the messy theory with an infinite
number of tensor fields for future work.

On the other hand, the noncommutativity at the boundary of
a membrane in the $C$-field background in M-theory is the loop-space
noncommutativity $[x^\mu(\sigma),x^\nu(\sigma')]\sim i\epsilon^{\mu\nu\rho}
x_\rho'(\sigma) \delta(\sigma-\sigma')$ \cite{berg,kawa}.
This noncommutativity obviously respects the translational symmetry.
Since the noncommutative string theory comes from a limit of
M-theory, the theory might be controllable.
Thus, concerning the interests in noncommutative field theories
in more than two-dimensions, the noncommutative string theory
might be interesting.

We have finally discussed an analogue of thermodynamics of 
noncommutative free scalar field theory. What we have computed is the
partition function of free scalar field theory in a noncommutative
three-dimensional space with Euclidean signature, following the 
common trick in usual commutative field theories.
The result shows the reduction of the degrees of freedom
in the ultraviolet, which comes essentially from the compactness of
the momentum space in the case of Euclidean signature.
However it is not clear to us whether the partition function is
really related to the thermodynamics of the noncommutative field
theory, since the noncommutativity of the time coordinate makes
it hard to define statistical thermodynamics in a convincing way.

There seem to exist several puzzles on the three-dimensional
noncommutative field theory such as unitarity and renormalizability.
To analyse these issues perturbatively, we need to develop further the 
computational technic of $SL(2,R)$ group theoretic Feynman integral.
We hope that the present work may be meaningful as a primitive step toward
consistent treatment of quantum field theories 
in more than two-dimensional noncommutative space-times.

\vspace{.5cm}
\noindent
{\large\bf Acknowledgements}\\[.2cm]
The authors would like to thank N.\ Ikeda for valuable discussions.
N.S. is supported in part by Grant-in-Aid for Scientific Research
(\#12740150), and in part by Priority Area:
``Supersymmetry and Unified Theory of Elementary Particles'' (\#707),
from Ministry of Education, Science, Sports and Culture.

\newcommand{\J}[4]{{\sl #1} {\bf #2} (#3) #4}
\newcommand{\andJ}[3]{{\bf #1} (#2) #3}
\newcommand{\AP}{Ann.\ Phys.\ (N.Y.)}
\newcommand{\MPL}{Mod.\ Phys.\ Lett.}
\newcommand{\IJMP}{Int.\ J.\ Mod.\ Phys.}
\newcommand{\NP}{Nucl.\ Phys.}
\newcommand{\PL}{Phys.\ Lett.}
\newcommand{\PR}{Phys.\ Rev.}
\newcommand{\PRL}{Phys.\ Rev.\ Lett.}
\newcommand{\PTP}{Prog.\ Theor.\ Phys.}
\newcommand{\JMP}{J.\ Math.\ Phys.}
\newcommand{\th}{hep-th/}
\newcommand{\CMP}{Comm.\ Math.\ Phys.}

\end{document}